\documentclass[twocolumn,prx,preprintnumbers,superscriptaddress,amsmath,amssymb,english]{revtex4}
\usepackage{amsmath,amssymb,amsfonts,mathrsfs}
\usepackage{amsthm}
\usepackage{CJK}
\usepackage{bm}
\usepackage{babel}
\usepackage{graphicx}
\allowdisplaybreaks
\usepackage{braket}
\usepackage[breaklinks]{hyperref}
\usepackage{color}
\hypersetup{colorlinks=true, linkcolor=blue, citecolor=blue, filecolor=blue, urlcolor=blue}
\UseRawInputEncoding

\begin{document}
\title{Electric and magnetic fields tuned spin-polarized topological phases in two-dimensional ferromagnetic MnBi$_4$Te$_7$}

\author{Shi Xiao}
\altaffiliation{S. X. and X. L. X. contributed equally to this work.}
\affiliation{Institute for Structure and Function and Department of Physics, Chongqing University, Chongqing 400044, P. R. China}
\affiliation{Chongqing Key Laboratory for Strongly Coupled Physics, Chongqing 400044, P. R. China}

\author{Xiaoliang. Xiao}
\altaffiliation{S. X. and X. L. X. contributed equally to this work.}
\affiliation{Institute for Structure and Function and Department of Physics, Chongqing University, Chongqing 400044, P. R. China}
\affiliation{Chongqing Key Laboratory for Strongly Coupled Physics, Chongqing 400044, P. R. China}

\author{Fangyang Zhan}
\affiliation{Institute for Structure and Function and Department of Physics, Chongqing University, Chongqing 400044, P. R. China}
\affiliation{Chongqing Key Laboratory for Strongly Coupled Physics, Chongqing 400044, P. R. China}

\author{Jing Fan}
\affiliation{Center for Computational Science and Engineering, Southern University of Science and Technology, Shenzhen 518055, P. R. China}

\author{Xiaozhi Wu}
\affiliation{Institute for Structure and Function and Department of Physics, Chongqing University, Chongqing 400044, P. R. China}
\affiliation{Chongqing Key Laboratory for Strongly Coupled Physics, Chongqing 400044, P. R. China}

\author{Rui Wang}
\email[]{rcwang@cqu.edu.cn}
\affiliation{Institute for Structure and Function and Department of Physics, Chongqing University, Chongqing 400044, P. R. China}
\affiliation{Chongqing Key Laboratory for Strongly Coupled Physics, Chongqing 400044, P. R. China}
\affiliation{Center for Computational Science and Engineering, Southern University of Science and Technology, Shenzhen 518055, P. R. China}
\affiliation{Center for Quantum Materials and Devices, Chongqing University, Chongqing 400044, P. R. China}

\date{\today}

\begin{abstract}
Applying electric or magnetic fields is widely used to not only  create and manipulate topological states but also facilitate their observations in experiments. In this work, we show by first-principles calculations and topological analysis that the time-reversal (TR) symmetry-broken quantum spin Hall (QSH) state emerges in a two-dimensional (2D) ferromagnetic  MnBi$_4$Te$_7$ monolayer. This TR-symmetry broken QSH phase possesses a highly tunable nontrivial band gap under an external electric field (or tuning interlayer distance). Furthermore, based on the Wannier-function-based tight-binding approach, we reveal that a topological phase transition from the TR-symmetry broken QSH phase to the quantum anomalous Hall (QAH) phase occurs with the increase of magnetic fields. Besides, we also find that a reverse electric fields can facilitate the realization of QAH phase. Our work not only uncovers the ferromagnetic topological properties the MnBi$_4$Te$_7$ monolayer tuned by electric and magnetic fields, but also can stimulate further applications to spintronics and topological devices.
\end{abstract}

\maketitle
\section{introduction}

Over the past decades \cite{Haldane,TKNN,Hasan,QiZhang}, the exploration of topological phases of matter has attracted intensive interest in condensed matter physics. Among various topological phases, one of the most important members is the time-reversal (TR) invariant topological insulator (TI) with a $\mathbb{Z}_2$ index, which was first predicted in two-dimensional (2D) graphene \cite{KaneMele1,KaneMele2} and HgTe quantum well \cite{SCZhang}, and experimentally confirmed soon in the latter \cite{TIexp}. The 2D TI with the $\mathcal{T}$-symmetry is also termed as the quantum spin Hall (QSH) insulators. Later for a long time, the TR-symmetry is considered to be a prerequisite to QSH states.
In fact, when the $\mathcal{T}$-symmetry is absent, the combination of topology and magnetism can lead to more fruitful spin-polarized nontrivial phases, such as quantum anomalous Hall (QAH) effects \cite{QAHreal1,QAHreal2,QAHreal3,QAHreal4,QAHreal5,QAHreal6}, TR-symmetry broken QSH effects \cite{TRSBQSH}, topological magnetoelectric effects \cite{Magnetoelectric,Magnetoelectric1}, Majorana fermions \cite{Majorana,Majorana1,Majorana2,Majorana3,Majorana4}, and magnetic Weyl fermions \cite{TWS,Chan}. Among these topological phenomena, the QAH effect with dissipationless chiral edges and quantized Hall conductance is one of the most fascinating topics.

The QAH effect was first been observed experimentally in magnetically doped TI thin films \cite{QAHreal1}. However, magnetic dopants require that the  QAH effect only occurs in ultralow temperatures due to inhomogeneity of ferromagnetic order. Comparing with magnetically doped systems, an undoped stoichiometric material prefers to form ordered magnetic structures. Therefore, for purposes of practical application, there have been many theoretical proposals for realizing high-temperature QAH effects in intrinsical magnetic insulating systems \cite{HTIQAH,HTIQAH1,HTIQAH2,HTIQAH3,HTIQAH4,HTIQAH5,HTIQAH6}. But, unfortunately, the experimental confirmation of QAH effects in intrinsic magnetic materials has been relatively slow.

Recently, the van der Waals (vdW) Mn-Bi-Te superlattices MnBi$_2$Te$_4$/(Bi$_2$Te$_3$)$_n$ ($n=0,1,2,3,\ldots$) have guided significant advancements of QAH effects or other spin-polarized topological phases with intrinsic mechanism both in theories and experiments \cite{TE,TE1,QAHreal3,QAHreal4}. The compound MnBi$_2$Te$_4$ (i.e., $n=0$) is composed of Te-Bi-Te-Mn-Te-Bi-Te septuple layers, and compounds MnBi$_2$Te$_4$/(Bi$_2$Te$_3$)$_n$ ($n=0,1,2,3,\ldots$) are in essence $1:n$ formalism of a MnBi$_2$Te$_4$ septuple layer and a Bi$_2$Te$_3$ quintuple layer. These layered compounds exhibit intralayer ferromagnetism and interlayer antiferromagnetism, resulting in their topological features dependent on odd or even numbers of layers \cite{MnBiTe2,MnBiTe3,MnBiTe4,MnBiTe13}. Remarkably, the QAH effect up to 1.4 K without the aid of external magnetic fields has been observed in a five-septuple-layer of MnBi$_2$Te$_4$ \cite{QAHreal3}. Beyond MnBi$_2$Te$_4$, other vdW Mn-Bi-Te superlattices such as MnBi$_4$Te$_7$ and MnBi$_6$Te$_{10}$ (i.e., $n=1, 2$) also currently stimulate extensive interest to realize various spin-polarized topological phases \cite{QAHreal4,MnBiTe1,MnBiTe5,MnBiTe6,MnBiTe7,MnBiTe8,MnBiTe9,MnBiTe10,MnBiTe11,MnBiTe12}.

\begin{figure*}[htb]
    \centering
    \includegraphics[width=0.89\textwidth]{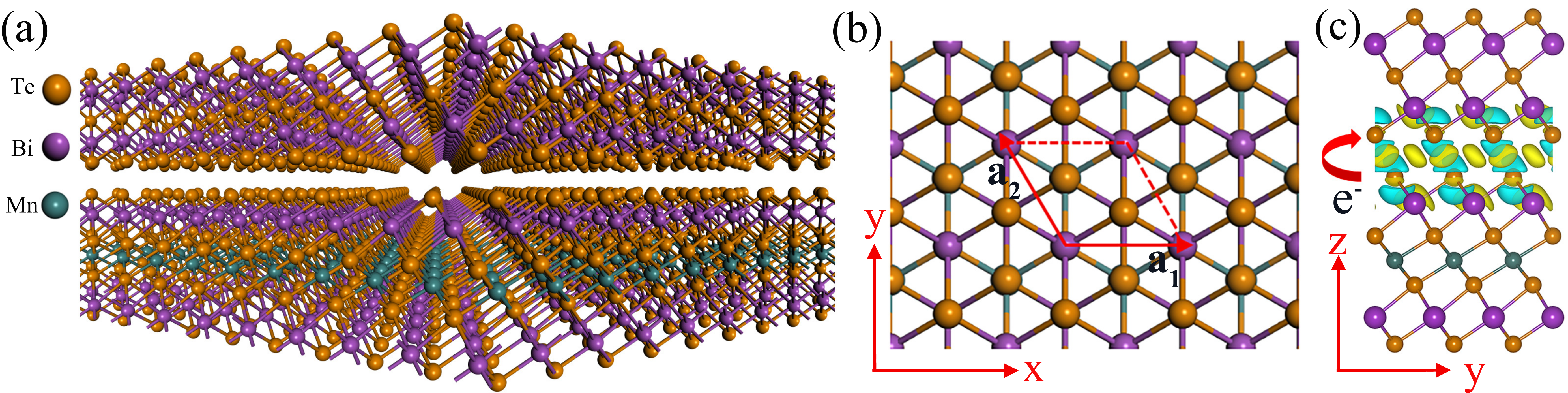}%
    \caption{
    (a) Side and (b) top views of the Bi$_2$Te$_3$/MnBi$_2$Te$_4$ heterostructure (i.e., the MnBi$_4$Te$_7$ monolayer). In panel (b), the hexagonal unit cell is marked by red-lines and two basic vectors $\mathbf{a}_1$ and ${\mathbf{a}_2}$ satisfy $|\mathbf{a}_1|=|\mathbf{a}_1|=a$. (c) The differential charge density (isosurface value of 0.0001$e$/Bohr$^3$), which indicates that electrons transfer from the MnBi$_2$Te$_4$ layer to the Bi$_2$Te$_3$ layer. Yellow and blue isosurface respectively denote electrons accumulation and reduction.}
    \label{fig1}
\end{figure*}

As is well known, applying an external field such as electric or magnetic fields is widely used to not only tune topological states but also facilitate their observations in experiments \cite{QAHreal3,TRSBQSH,Efield1,Efield2,Efield3,Efield4,Efield5,Efield,Bfield,Bfield1,Bfield2,Bfield3}. For instance, the observed temperature of QAH effects in the five-septuple-layer MnBi$_2$Te$_4$ film under a small magnetic field can remarkably be increased up to 6.5 K \cite{QAHreal3}, and the combination of a magnetic field and back gate voltage can give rise to a high-Chern-number QAH effect in ten-septuple-layer MnBi$_2$Te$_4$ film \cite{MnBiTe14} or drive a phase transition from the axion insulating phase to the QAH phase in the six-septuple-layer MnBi$_2$Te$_4$ film \cite{Bfield3,MnBiTe15} at relatively high temperatures. Therefore, a theoretical study of spin-polarized topological phases in Mn-Bi-Te systems tuned by external electric or magnetic fields is desirable.

In this work, based on first-principles calculations and topological analysis, we identify that the MnBi$_4$Te$_7$ monolayer with a ferromagnetic (FM) groundstate exhibits the TR-symmetry broken QSH effect, which possesses a quantized spin Hall conductance associated with the nonzero spin Chern number. The nontrivial band gap of the TR-symmetry broken QSH phase can be considerably enlarged by applying an external positive electric fields perpendicular to the MnBi$_4$Te$_7$ atomic plane.
Furthermore, using the Wannier-function-based tight-binding (WFTB) approach, we reveal that a topological phase transition from the TR-symmetry broken QSH phase to the QAH phase occurs with the increase of Zeemann exchange energy of magnetic fields. Finally, we present a phase diagram to fully understand cooperative effects of electric and magnetic fields.

\section{Computational methods}

To investigate electronic properties of monolayer MnBi$_4$Te$_7$, we carried out first-principles calculations as implemented in Vienna \textit{ab initio} simulation package \cite{Kresse1,Kresse2} based on the density-functional theory (DFT) \cite{Kohn1,Kohn2}. The generalized gradient approximation (GGA) with Perdew-Burke-Ernzerhof (PBE) formalism was employed to describe the exchange-correlation functional \cite{PBE}. The electron-ion interaction was treated by projector-augmented-wave potentials \cite{PAW}. The cutoff energy of the plane wave basis was set to 500 eV, and the Brillouin zone was sampled with a $12\times12\times1$ Monkhorst-Pack grid \cite{BZ}. Since the correlation effects of Mn $3d$ electrons, we employed the GGA+U method \cite{GGAU} and set the $U=5$ eV \cite{U5,U5_1}. The forces on each atom were relaxed to be less than 0.001 eV/{\AA}.  A 15 {\AA} vacuum layer along the $z$ direction was used to avoid the interaction between neighboring layers. The effects of applying external electric fields normal to the MnBi$_4$Te$_7$ monolayer was directly including in DFT calculations. The topological properties were revealed by constructing a  WFTB Hamiltonian based on maximally localized Wannier functions methods combining DFT calculations \cite{Wannier90,Wannier901,Wannier902}. The WFTB Hamiltonian with Mn $d$, Bi $p$, and Te $p$ orbitals can well reproduce the DFT electronic band structures including spin-orbital coupling (SOC) effects in the range of $E_F \pm 2$ eV [see Fig. S1]. The topological edge states such as the local density of states (LDOS) were computed using WANNIERTOOLS package based on the iterative Greens method \cite{WannierTools,Green}.

To reveal the electronic properties under magnetic fields, we consider an Zeeman exchange energy
\begin{equation}
H_{\rm Z} = \tilde{g} \mu_B (\mathbf{L}+2\mathbf{S}) \cdot \mathbf{B},
\end{equation}
in which $\tilde{g}$, $\mu_B$, $\mathbf{L}$, and $\mathbf{S}$ represent the effective $g$-factor, Bohr magneton, orbital angular momentum, and spin angular momentum, respectively. As a result, the total Hamiltonian can be expressed as $H_T=H_{\rm SOC}(\mathbf{k})+H_{\rm Z}$, in which the WFTB Hamiltonian $H_{\rm SOC}(\mathbf{k})$ in the presence of SOC is
\begin{equation}
\begin{split}
H_{\rm SOC}(\mathbf{k}) &= \sum_{\alpha\beta}\big< \psi_{\mathbf{k},\alpha} | H_{\rm SOC} | \psi_{\mathbf{k},\beta} \big> \\
&= \sum_{\alpha\beta} \sum_{\mathbf{R}} e^{i \mathbf{k} \cdot \mathbf{R}} t_{\alpha\beta} (\mathbf{R} - 0) \\
\end{split}
\end{equation}
with
\begin{equation}
t_{\alpha\beta} (\mathbf{R} - 0) = \big< \mathbf{0} + \mathbf{s}_{\alpha} | H_{\rm SOC} | \mathbf{R} + \mathbf{s}_{\beta} \big>.
\end{equation}
Here, $| \psi_{\mathbf{k}} \big>$ are Bloch states over the $\mathbf{k}$ space, $\mathbf{R}$ is the Bravais lattice vector, and $t_{\alpha\beta} (\mathbf{R} - 0)$ are the hopping parameters from orbital $\beta$ at site $\mathbf{s}_{\beta}$ in the home cell at $\mathbf{R}=0$ to orbital $\alpha$ at site $\mathbf{s}_{\alpha}$ in the unit cell located at $\mathbf{R}$.  Note that we here ignore the Peierls phase in the hopping parameters $t_{\alpha\beta}$ or Landau levels. This approach is reliable and effective for studying the magnetic field induced evolution of topological states \cite{PhysRevB.98.075123,PhysRevB.101.035105}.

\section{Results and Discussion}

As shown in Fig. \ref{fig1}, the MnBi$_4$Te$_7$ monolayer can be considered as the Bi$_2$Te$_3$/MnBi$_2$Te$_4$ heterostructure with the hexagonal symmetry, i.e., a Bi$_2$Te$_3$ quintuple layer on top of a MnBi$_2$Te$_4$ septuple layer. The hexagonal unit cell is denoted by two basic vectors $\mathbf{a}_1$ and ${\mathbf{a}_2}$, which satisfy  $|\mathbf{a}_1|=|\mathbf{a}_1|=a$ [see Fig. \ref{fig1}(b)]. The optimized in-plane lattice constants are 4.37 {\AA} and 4.43 {\AA} for MnBi$_2$Te$_4$ and Bi$_2$Te$_3$, respectively, and thus the small difference ($\sim 1.35\%$) of their lattice constants only leads to a tiny mismatch between two layers. The optimized equilibrium distance between Bi$_2$Te$_3$ and MnBi$_2$Te$_4$ layers is 2.68 {\AA}, indicating that there is a considerable interlayer hybridization between Bi$_2$Te$_3$ and MnBi$_2$Te$_4$ layers. As shown in Fig. \ref{fig1}(c), the differential charge density indicates that electrons transfer from the MnBi$_2$Te$_4$ layer to the Bi$_2$Te$_3$ layer, forming a built-in electric field in MnBi$_4$Te$_7$. In this case, outer shell electrons of Bi and Te atoms near the interface would suffer a charge redistribution to maintain the charge balance. As a result, though the isolated Bi$_2$Te$_3$ and MnBi$_2$Te$_4$ layers are both trivial, their interlayer coupling [see Fig. \ref{fig1}(c)] can facilitate the formation of band inversion and nontrivial band topology in MnBi$_4$Te$_7$. Bulk Bi$_2$Te$_3$ and MnBi$_2$Te$_4$ are both typical vdW layered materials; that is, their monolayers are easily exfoliated in experiments. The binding energy is calculated by $E_b= E_{\mathrm{Bi_2Te_3}}+E_{\mathrm{MnBi_2Te_4}}-E_{\mathrm{MnBi_4Te_7}}$, where $E_{\mathrm{Bi_2Te_3}}$, $E_{\mathrm{MnBi_2Te_4}}$, and $E_{\mathrm{MnBi_4Te_7}}$ are the total energies of the isolated Bi$_2$Te$_3$ layer, isolated MnBi$_2$Te$_4$ layer, and MnBi$_4$Te$_7$ monolayer, respectively. The calculated $E_b= 380$ meV, implying that the formation of the  Bi$_2$Te$_3$/MnBi$_2$Te$_4$ (i.e, $\mathrm{MnBi_4Te_7}$)  heterostructure remarkably improves the thermodynamic stability.

\begin{figure}[htb]
    \centering
    \includegraphics[width=0.49\textwidth]{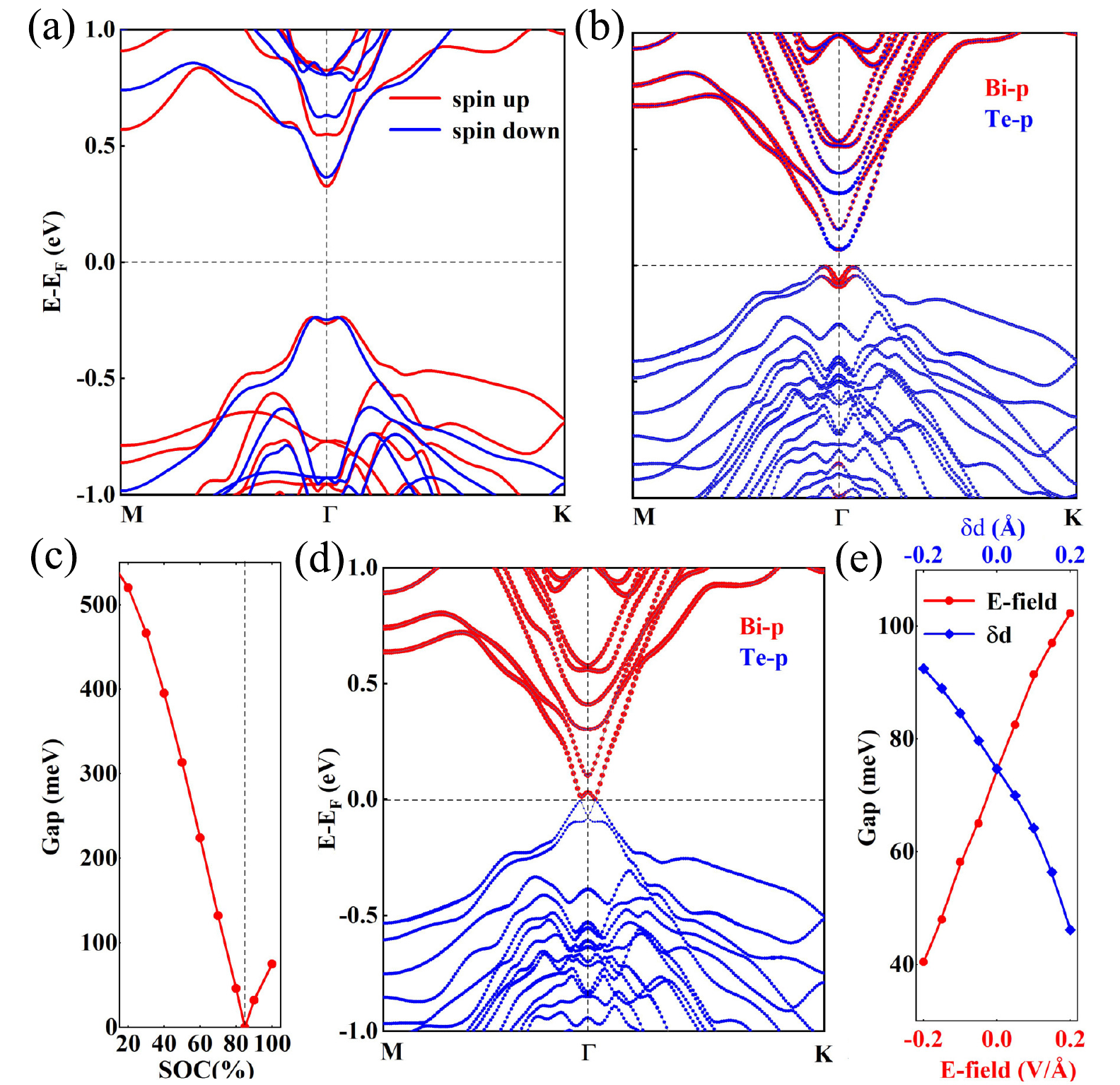}%
    \caption{
    (a) The spin-polarized electronic band structure of $\mathrm{MnBi_4Te_7}$ along high-symmetry paths in the absence of SOC. The spin-up and spin-down bands are denoted by red and blue lines, respectively. (b) Orbital-resolved electronic band structures of $\mathrm{MnBi_4Te_7}$ along high-symmetry paths in the presence SOC. The component of Bi $p$ (Te $p$) orbitals is proportional to
the width of the red (blue) curves. (c) The band gap of $\mathrm{MnBi_4Te_7}$ as a function of the SOC strength. The system would undergo a topological phase transition with band gap closing and then reopening if gradually turning on SOC. (d) Orbital-resolved electronic band structures of $\mathrm{MnBi_4Te_7}$ along high-symmetry paths at the critical value of SOC strength ($\sim$85\% of intrinsic SOC strength). (e) The band gap as a function of the electric field strength $E$ (or the variation of interlayer distance $\delta d$).}
    \label{fig2}
\end{figure}

Our DFT calculations on the electronic properties indicate that the $\mathrm{MnBi_4Te_7}$ monolayer possesses a FM groundstate with a saturation magnetic moment of $\sim$5.00 $\mu_{B}$, which matches well with prior results \cite{MnBiTe2,MnBiTe16}. The spin-polarized electronic band structure of $\mathrm{MnBi_4Te_7}$ in the absence of SOC is shown in Fig. \ref{fig2}(a). We can see that the conduction and valance bands at the $\Gamma$ point are respectively contributed by spin-up and spin-down channels. Meanwhile, conduction band minimum (CBM) is located at the $\Gamma$ point and valence band maximum (VBM) is slightly away from the $\Gamma$ point. As a result, the $\mathrm{MnBi_4Te_7}$ monolayer is a FM semiconductor. 

Within SOC, the coupling between the spin and orbital sectors leads to spontaneous magnetization. Hence, we performed total-energy calculations with possible magnetic configurations to determine the magnetization direction. The results confirm that the easy axis along the out-of-plane direction is perpendicular to the $\mathrm{MnBi_4Te_7}$ plane. The electronic band structures of $\mathrm{MnBi_4Te_7}$ in the presence of SOC are depicted in Fig. \ref{fig2}(b). As expected, due to the strong SOC strength of Bi element, it is found that the SOC effects drastically modify band profiles and especially for the band gap.
We also illustrate orbital-resolved contributions in Fig. \ref{fig2}(b). The figure shows that a visible band inversion at the $\Gamma$ point. This band inversion happens between Bi $p$ and Te $p$ orbitals, which is essentially the same as Bi$_2$Te$_3$ \cite{Bi2Te3}. To reveal the band-reversal process, we plot the band gap at the $\Gamma$ point as a function of the SOC strength in Fig. \ref{fig2}(c). One can see that the band gap first closes and then reopens. At the critical gapless point, the bands composed of Bi $p$ and Te $p$ orbitals touch at the Fermi level [see Fig. \ref{fig2}(d)]. Therefore, beyond the threshold of SOC strength ($\sim$85\% of intrinsic SOC strength), a topological phase transition occurs, meaning that the $\mathrm{MnBi_4Te_7}$ monolayer is a topologically nontrivial insulating phase with the inverted band gap of $\sim$75 meV.

Since the built-in electric field induced by interlayer hybridization and charge transfer is present between Bi$_2$Te$_3$ and MnBi$_2$Te$_4$ layers, the nontrivial band gap of $\mathrm{MnBi_4Te_7}$ can be tuned by applying external electric fields (or changing interlayer distance). As shown in Fig. \ref{fig2}(e), we schematically illustrate the band gap as a function of  $E$ or $\delta d$. Here, the electric field $E$ with its positive direction along $z$ axis is normal to the MnBi$_4$Te$_7$ plane; meanwhile, the variation of interlayer distance $\delta d$ is defined as $\delta d =d-d_0$ with the optimized equilibrium distance $d_0$. As expected,  it is found that a positive electric field can effectively enhance the interlayer hybridization and thus enlarge the nontrivial band gap. As shown in Fig. S2 in the SM \cite{SM}, we can see that the band inversion at $\Gamma$ is still preserved when the electric field strength increases to 0.2 V/{\AA}. In this case, the nontrivial band gap can be considerable up to $\sim$102 meV [see Fig. \ref{fig2}(e)]. That is to say, applying external electric fields or changing interlayer distance in the physical regime does not change the nontrivial band topology of MnBi$_4$Te$_7$.

\begin{figure}[htb]
    \centering
    \includegraphics[width=0.47\textwidth]{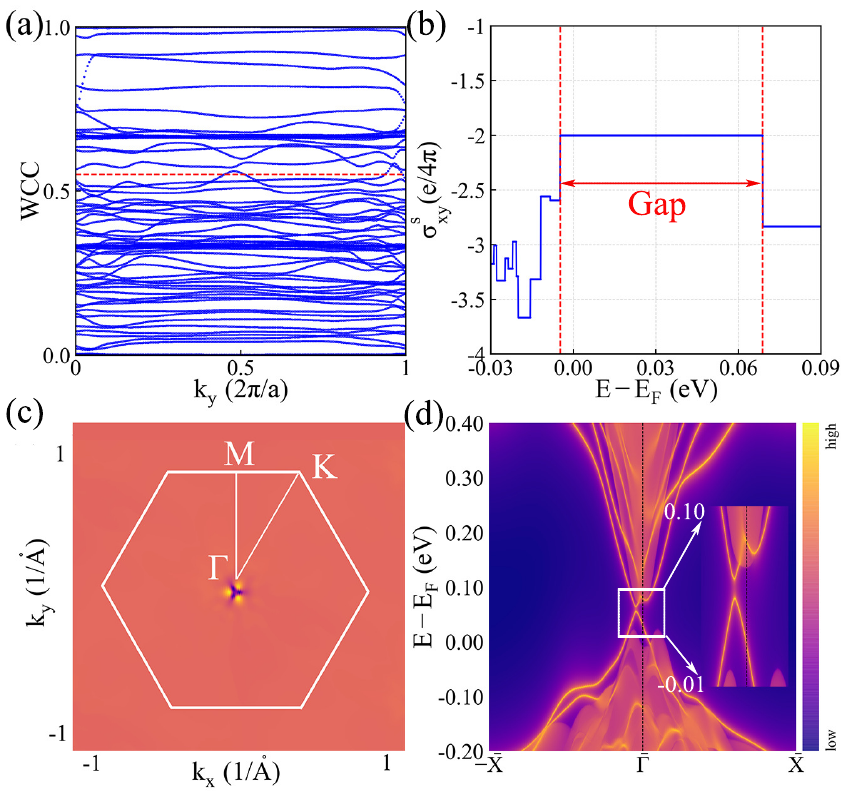}%
    \caption{
    (a) The evolution of WCC of MnBi$_4$Te$_7$ in the full closed plane of $k_z = 0$. The referenced line is denoted by the red dashed line. (b) The spin Hall conductance $\sigma_{xy}^s$ of MnBi$_4$Te$_7$ as a function of the Fermi level. Inside the bang gap, the spin Hall conductance $\sigma_{xy}^s$ is quantized to $C^s\frac{e}{4\pi}$ with  $C^s = 2$. (c) The Berry curvature distribution of MnBi$_4$Te$_7$. The first BZ is marked by the white lines. (d) The calculated LDOS of a semi-infinite of MnBi$_4$Te$_7$. The inset shows that the Dirac cone formed from two edge states is slightly gapped.
    }
    \label{fig3}
\end{figure}

For a FM nontrivial system, there are usually two distinct topological phases with bulk band gaps \cite{TRSBQSH}. The one is the TR-symmetry broken QSH insulator with nonzero spin Chern number $\mathcal{C}^s$, and the other is the QAH insulator with nonzero Chern number $\mathcal{C}$. The Chern number $\mathcal{C}$ and spin-Chern number $\mathcal{C}^s$ are respectively defined as $\mathcal{C} = \mathcal{C}_{\uparrow}+\mathcal{C}_{\downarrow}$ and $\mathcal{C}^s = \mathcal{C}_{\uparrow}-\mathcal{C}_{\downarrow}$, where $\mathcal{C}_{\uparrow}$ and $\mathcal{C}_{\downarrow}$ are the spin-dependent Chern numbers of spin-up and spin-down sectors, respectively. To determine the topological phase of MnBi$_4$Te$_7$ monolayer, we calculate the evolution of Wannier charge centers (WCC) for the full closed plane of $k_z = 0$ using the Wilson loop method \cite{WCC}. As shown in Fig. \ref{fig3}(a), the calculated results indicate that the referenced line can always cross the the WCC line even times, i.e., the Chern number $\mathcal{C}=0$. Hence, the MnBi$_4$Te$_7$ monolayer does not belong to the QAH insulator but may be the TR-symmetry broken QSH insulator. To further confirm this, we calculated the spin Hall conductance $\sigma_{xy}^s$ using the Kubo formula \cite{Kubo},
\begin{equation}\label{eqc}
\begin{split}
\sigma_{xy}^s & =e\hbar \int {\frac{dk_x dk_y}{(2\pi)^2}}\Omega^{s}_{\mathbf{k}}, \\
\Omega^{s}_{\mathbf{k}} &= -2\text{Im} \sum_{m \neq n} \frac{ \big< \psi_{n\mathbf{k}} | \mathbf{j}_{x}^s | \psi_{m\mathbf{k}} \big> \big< \psi_{m\mathbf{k}} | \mathbf{v}_{y} | \psi_{n\mathbf{k}} \big>}{(E_{m}-E_{n})^{2}},
\end{split}
\end{equation}
where $\mathbf{v}_i$ ($i=x,y$) is the velocity operator, $E_{n\mathbf{k}}$ is the eigenvalue of the Bloch functions $\psi_{n\mathbf{k}}$, and $\mathbf{j}_x^s=1/2\{\mathbf{s}^z,\mathbf{v}_x\}$ denotes a spin current operator in the $x$ direction. From Eq. (\ref{eqc}), we calculate the spin Hall conductance $\sigma_{xy}^S$ with WFTB Hamiltonian using WANNIER90 package \cite{Wannier901}. The calculated $\sigma_{xy}^s$ of MnBi$_4$Te$_7$ as a function of the Fermi level is plotted in Fig. \ref{fig3}(b). We can see that the value of $\sigma_{xy}^s$ is quantized to $C^s\frac{e}{4\pi}$ with  $C^s = -2$ inside the nontrivial band gap, which is contributed from the spin Berry curvature $\Omega^{s}_{\mathbf{k}}$ in the momentum space. As shown in Fig. \ref{fig3}(c), we can see that both the positive and negative $\Omega^{s}_{\mathbf{k}}$, which are with respective to the three-fold rotational symmetry, clearly diverges near the $\Gamma$ point, uncovering the nontrivial band topology of  MnBi$_4$Te$_7$. We show the calculated LDOS of a semi-infinite of MnBi$_4$Te$_7$ in Fig. \ref{fig3}(d). Distinct from the gapless Dirac cone of TR-symmetry protected QSH phase, the Dirac cone formed from two edge states is slightly gapped [see the inset of Fig. \ref{fig3}(d)], demonstrating the presence of TR-symmetry broken QSH phase in MnBi$_4$Te$_7$ \cite{TRSBQSH}.

Compared with the TR-symmetry broken QSH phase, the QAH phase with chiral edge states and quantized anomalous Hall conductance is more interesting. The previous investigation demonstrated that the competition between the QAH and TR-symmetry broken QSH phases was dependent on the relation of Rashba SOC and Zeeman exchange fields \cite{TRSBQSH}. On the one hand, applying external electric fields (or changing interlayer distance) can potentially affect the Rashba SOC and then tune the nontrivial band gap of TR-symmetry broken QSH phase [see Fig. \ref{fig2}(e)]. On the other hand, an external magnetic field can effectively enhance the magnetic exchange energy in intrinsically magnetic material. Therefore, we next study the evolution of topological properties in MnBi$_4$Te$_7$ under external magnetic fields.

\begin{figure}[htb]
    \centering
    \includegraphics[width=0.47\textwidth]{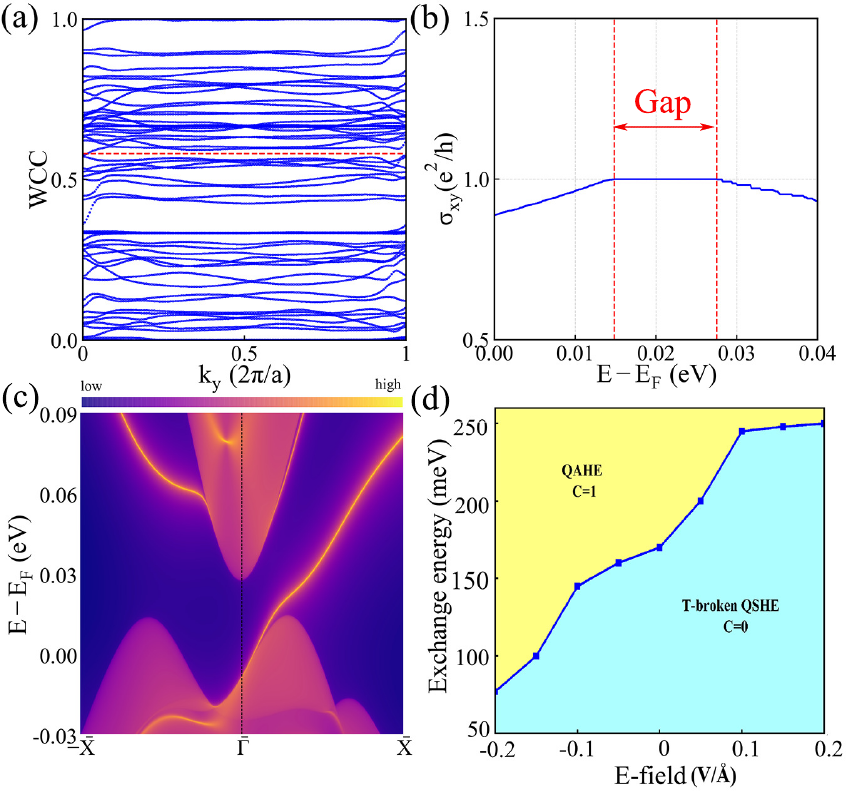}%
    \caption{(a) The evolution of WCC of MnBi$_4$Te$_7$ in the full closed plane of $k_z = 0$. The referenced line is denoted by the red dashed line. Here, the perpendicular electric field strength and Zeeman exchange energy are respectively set to -0.2 V/{\AA} and 130 meV. In this case the QAH phase is present. (b) The intrinsic Hall conductance $\sigma_{xy}$ of MnBi$_4$Te$_7$ as a function of the Fermi level. Inside the bang gap, $\sigma_{xy}$ is quantized to $e^2/h$. (c)  The calculated LDOS exhibits chiral edge states of QAH phase. (d) The phase diagram as functions of the Zeeman exchange energy and electric field strength.}
    \label{fig4}
\end{figure}

By applying an external magnetic fields $\mathbf{B}$ normal to the MnBi$_4$Te$_7$ plane along the $z$ axis, one can find that the band gap first closes and then reopens (see Fig. S3 in the SM \cite{SM}), implying the occurrence of a topological phase transition. The critical gapless point at the $K$ valley occurs at the Zeeman exchange energy of 170 meV. Beyond the threshold of Zeeman exchange energy, we calculate the evolution of WCC for the full closed plane of $k_z = 0$ as shown in Fig. \ref{fig4}(a).  The referenced line can always cross the WCC line one times, implying the presence of QAH phase with $\mathcal{C}=1$. To further confirm that the QAH phase is indeed present in MnBi$_4$Te$_7$, we also calculate the intrinsic Hall conductance $\sigma_{xy}$ [see Fig. \ref{fig4}(b)]. It is found that the quantized plateau of $\mathcal{C}\frac{e^{2}}{h}$ with $\mathcal{C}=1$ inside the nontrivial band gap is clearly visible. The semi-infinite LDOS of MnBi$_4$Te$_7$ is plotted in Fig. \ref{fig4}(c), exhibiting one nontrivial chiral edge state connecting the valence and conduction bands. As depicted in Fig. \ref{fig2}(e), a reverse electric field can effectively weaken the Rashba SOC and thus reduce the nontrivial band gap of TR-symmetry broken QSH phase, which is favourable to realize topological phase transition from the TR-symmetry broken QSH phase to the QAH phase. Therefore, to fully understand electric and magnetic fields tuned spin-polarized topological properties of MnBi$_4$Te$_7$, we calculate a phase diagram as functions of the Zeeman exchange energy and electric field strength [see Fig. \ref{fig4}(d)].  We can see that the electric field can effectively reduce the threshold of Zeeman exchange energy, thus facilitating to experimentally achieve the QAH phase in MnBi$_4$Te$_7$.

\section{Summary}
In summary, based on first-principles calculations and the WFTB approach, we uncover manipulated spin-polarized topological phases manipulated by the external electric and magnetic fields in the MnBi$_4$Te$_7$ monolayer with a FM groundstate. The external electric and magnetic fields respectively affect the Rashba SOC and magnetic exchange energy, and thus tune the nontrivial band gap and induce the topological phase transition. In the absence of external fields, the MnBi$_4$Te$_7$ monolayer possesses the  TR-symmetry broken QSH phase with the spin Chern number $\mathcal{C}^s = -2$. The nontrivial band gap of TR-symmetry broken QSH phase is highly tunable. Furthermore, the increase of Zeemann exchange energy of magnetic fields leads to a topological phase transition from the TR-symmetry broken QSH phase to the QAH phase with Chern number $\mathcal{C} = 1$. Our work can promote to further understand cooperative effects of electric and magnetic fields on spin-polarized topological phases. Besides, this findings can also stimulate potential applications of nontrivially topological spintronics in intensive Mn-Bi-Te systems.

\begin{acknowledgments}
This work was supported by the National Natural Science Foundation of China (NSFC, Grants No. 11974062, No. 12047564, and No. 11704177), the Chongqing Natural Science Foundation (Grants No. cstc2019jcyj-msxmX0563), the Fundamental Research Funds for the Central Universities of China (Grant No. 2020CDJQY-A057), and the Beijing National Laboratory for Condensed Matter Physics.
\end{acknowledgments}

%

\end{document}